\documentclass{aa} 
\pdfoutput=1

\usepackage{graphicx}
\usepackage{txfonts}
\begin{document}

   \title{FBQ 0951+2635: time delay and structure of the main lensing galaxy}    
                                                             
   \author{Vyacheslav N. Shalyapin\inst{1,2,3}
                \and
                Luis J. Goicoechea\inst{1,2}
                \and
                Eleana Ruiz-Hinojosa\inst{2}       
                }

\institute{Instituto de F\'isica de Cantabria (CSIC-UC), Avda. de Los Castros s/n, 
		E-39005 Santander, Spain\\
		\email{vshal@ukr.net, goicol@unican.es}
		\and
		Departamento de F\'\i sica Moderna, Universidad de Cantabria, 
                Avda. de Los Castros s/n, E-39005 Santander, Spain
                \and
		O.Ya. Usikov Institute for Radiophysics and Electronics, National 
                Academy of Sciences of Ukraine, 12 Acad. Proscury St., UA-61085 
                Kharkiv, Ukraine
                }

%   \date{Created in June 2024; ...}

% \abstract{}{}{}{}{} 
% 5 {} token are mandatory
 
\abstract{As there is a long-standing controversy over the time delay between the two images 
of the gravitationally lensed quasar FBQ 0951+2635, we combined early and new optical light 
curves to robustly measure a delay of 13.5 $\pm$ 1.6 d (1$\sigma$ interval). The new optical 
records covering the last 17 yr were also used to trace the long-timescale evolution of the 
microlensing variability. Additionally, the new time delay interval and a relatively rich 
set of further observational constraints allowed us to discuss the mass structure of the 
main lensing galaxy at redshift 0.26. This lens system is of particular interest because the
external shear from secondary gravitational deflectors is relatively low, but the external 
convergence is one of the highest known. When modelling the galaxy as a singular power-law 
ellipsoid without hypotheses/priors on the power-law index, ellipticity and position angle, 
we demonstrated that its mass profile is close to isothermal, and there is good alignment 
between mass and near-IR light. We also recovered the true mass scale of the galaxy. 
Finally, it is worth mentioning that a constant mass-to-light ratio model also worked 
acceptably well.}
   

\maketitle
%
%-------------------------------------------------------------------

\section{Introduction}
\label{sec:introd}

Strong gravitational lensing at the galaxy scale is a key tool to investigate the structure 
of non-local early-type galaxies \citep[e.g.,][]{2006ApJ...649..599K,
2024SSRv..220...87S}. If the lensed source is a quasar, the mass distribution of the main 
lensing galaxy can be probed with a suitable number of observational constraints. Although 
deep photometric observations with the $Hubble$ Space Telescope (HST) or ground-based 
facilities incorporating adaptive optics provide images of the quasar host galaxy 
\citep[e.g.,][]{2020MNRAS.498.1420W}, the basic constraints consist of the positions of the 
quasar images and the light distribution of the main lens. In addition to these basic 
astro-photometric data, deep spatially-resolved spectroscopy of the system is required to 
estimate unbiased macro-magnification ratios, i.e., flux ratios between quasar images that 
are not affected by extinction, microlensing, and intrinsic variability effects 
\citep[e.g.,][]{2012A&A...544A..62S,2016A&A...596A..77G}. Radio fluxes, if available, are 
expected to be unaffected by microlensing (large source) and dust extinction (long 
wavelength), and thus radio flux ratios at a single epoch are good proxies of the 
macro-magnification ratios for a lensed quasar with short delays between images. 
Sometimes, a smoothly-distributed lensing mass fails to reproduce measured 
macro-magnification ratios of a lensed quasar, and these flux ratio anomalies may be caused 
by the galaxy substructures \citep[millilensing; e.g.,][]{2001ApJ...563....9M}. Millilensing 
effects would be different for broad emission lines, radio fluxes and narrow emission lines 
because they depend on the source size \citep[e.g.,][]{2003MNRAS.339..607M}.  
     
According to Refsdal's 60-year old ideas \citep{1964MNRAS.128..307R}, the time delays 
between quasar images from photometric monitoring campaigns can also be used to constrain 
the mass distribution of the main lens, provided that spectroscopic redshifts of the source 
and main lens are known, and the main cosmological parameters are derived from independent 
experiments \citep[e.g.,][]{2009ApJS..180..225H,2020A&A...641A...6P}. Additionally,  
information about the stellar kinematics of the main lens, or about the secondary deflectors 
around the main one and along the line of sight is required to reliably probe the mass 
distribution of the main lensing galaxy \citep[e.g.,][]{1985ApJ...289L...1F,
2010ApJ...711..201S}. At present, this last information is only available for a relatively 
small number of lensed quasars with detailed studies on the stellar kinematics of their main 
lenses \citep[e.g.,][]{2020A&A...643A.165B,2023A&A...673A...9S}, or on main lens 
environments and line-of-sight deflectors \citep[e.g.,][]{2017MNRAS.467.4220R,
2017ApJ...850...94W}.

\object{FBQ 0951+2635} is a doubly imaged quasar that was discovered by 
\citet{1998AJ....115.1371S}. This lensed quasar is located at a redshift $z_{\rm{s}}$ = 
1.249 \citep{2015ApJS..219...29M}, and the main lens is an early-type galaxy at $z_{\rm{l}}$ 
= 0.260 \citep{2007A&A...465...51E}. It is a particularly interesting lens system for 
several reasons. First, even though near-IR HST imaging of \object{FBQ 0951+2635}
\citep{2000ApJ...543..131K} led to two formally precise solutions for the relative 
astrometry of the system and the morphology of the early-type galaxy 
\citep{2005A&A...431..103J,2012A&A...538A..99S}, these solutions are not consistent with 
each other. Thus, we aim to study how both astro-photometric datasets influence the 
reconstruction of the lensing mass. Although the near-IR morphology of the galaxy basically 
consists of a prominent elliptical halo, very recently, \citet{2023ApJ...952...54R} have 
suggested the existence of an additional feature: a faint edge-on disc. In this paper, we do 
not consider the presence of minor features requiring confirmation through new observations. 
Second, there are spectroscopic observations with large optical telescopes in very good 
seeing conditions and radio fluxes at 3.6 cm from the Very Large Array (VLA) that can be 
used to obtain a reliable constraint on the macro-magnification ratio 
\citep[e.g.,][]{2005A&A...431..103J,2012A&A...544A..62S}. Third, \citet{2017ApJ...850...94W} 
quantified the gravitational lensing effect of galaxy groups at the main lens redshift and 
along the line of sight.

We also aim to use the time delay between both quasar images as an important constraint on 
the mass distribution of the non-local early-type galaxy. However, current values of the 
time delay of \object{FBQ 0951+2635} are not trustworthy. Optical light curves in the period 
1999$-$2001 were initially analysed by \citet{2005A&A...431..103J}, who obtained a time 
delay of 16 $\pm$ 2 d (the brightest image is leading) using the last 38 data points. 
Despite these last data seem to be weakly affected by extrinsic (microlensing) variability, 
an iterative fit yielded a time delay of 13 $\pm$ 4 d using all (58) data points, i.e., 
having twice the uncertainty and a significantly shorter central value. From the same 
dataset, \citet{2011A&A...536A..44E} found that different time delay values in the 10$-$30 d 
interval are possible, whereas \citet{2015A&A...580A..38R} measured a time delay of 7.8 
$\pm$ 14.0 d. The obvious conclusion is that additional data are required to reliably 
measure the time delay of the system. 

The paper is organized as follows. In Sect.~\ref{sec:delay}, we analyse optical light curves 
of \object{FBQ 0951+2635} spanning 25 yr (1999$-$2024) to determine a reliable time delay 
between the two quasar images and identify the long-term microlensing signal. 
Sect.~\ref{sec:struc} includes the observational constraints for the lens system, as well as 
the methodology to probe the mass distribution of the main lensing galaxy and the results on 
its structure. In Sect.~\ref{sec:concl}, we present our main conclusions. 

\section{Optical light curves and time delay}
\label{sec:delay}

\subsection{Light curves}
\label{sec:records}

The first optical monitoring campaigns of \object{FBQ 0951+2635} focused on the $R$ 
passband. Thus, based on observations with the Nordic Optical Telescope (NOT) at 58 epochs 
(nights), \citet{2005A&A...431..103J} and \citet{2006A&A...455L...1P} presented $R$-band 
light curves of the two quasar images (A and B) from 1999 to 2001. A monitoring  programme 
with the main Maidanak Telescope (MT) in the period 2001$-$2006 also led to $R$-band light 
curves at 37 epochs \citep{2009MNRAS.397.1982S}. 

After the pioneering variability records from NOT and MT observations, as part of the 
GLENDAMA project \citep{2018A&A...616A.118G}, we have monitored \object{FBQ 0951+2635} with 
the Liverpool Telescope (LT) in the optical $r$ band from 2009 to 2024. Over the full 
monitoring period, $r$-band magnitudes at 142 epochs have been obtained \citep[the 
photometric model and initial LT light curves were presented by][]{2018A&A...616A.118G}. We 
then combined the LT magnitudes and those obtained with the Kaj Strand Telescope (KST) 
during 2008$-$2017 \citep{2023ApJ...952...54R}, removing two KST epochs at MJD = 54883 and 
57866 with possible outliers, correcting small magnitude offsets between the Tek2k and 
OneChip cameras in the KST (image A and B offsets of $-$0.001 and +0.005 mag for OneChip 
respect to Tek2k), and shifting the KST magnitudes to the LT system (+16.879 and +16.809 mag 
for A and B, respectively). We point out that magnitude offsets are estimated in a 
standard way by comparing concurrent data from two different cameras/telescopes. In 
addition, photometric errors rely on the intra-night magnitude scatter drawn from the 
analysis of two or three individual frames each night. Possible outliers are also removed as 
they may have an impact on time delay measurements. 

\begin{figure}
\centering
\includegraphics[width=9cm]{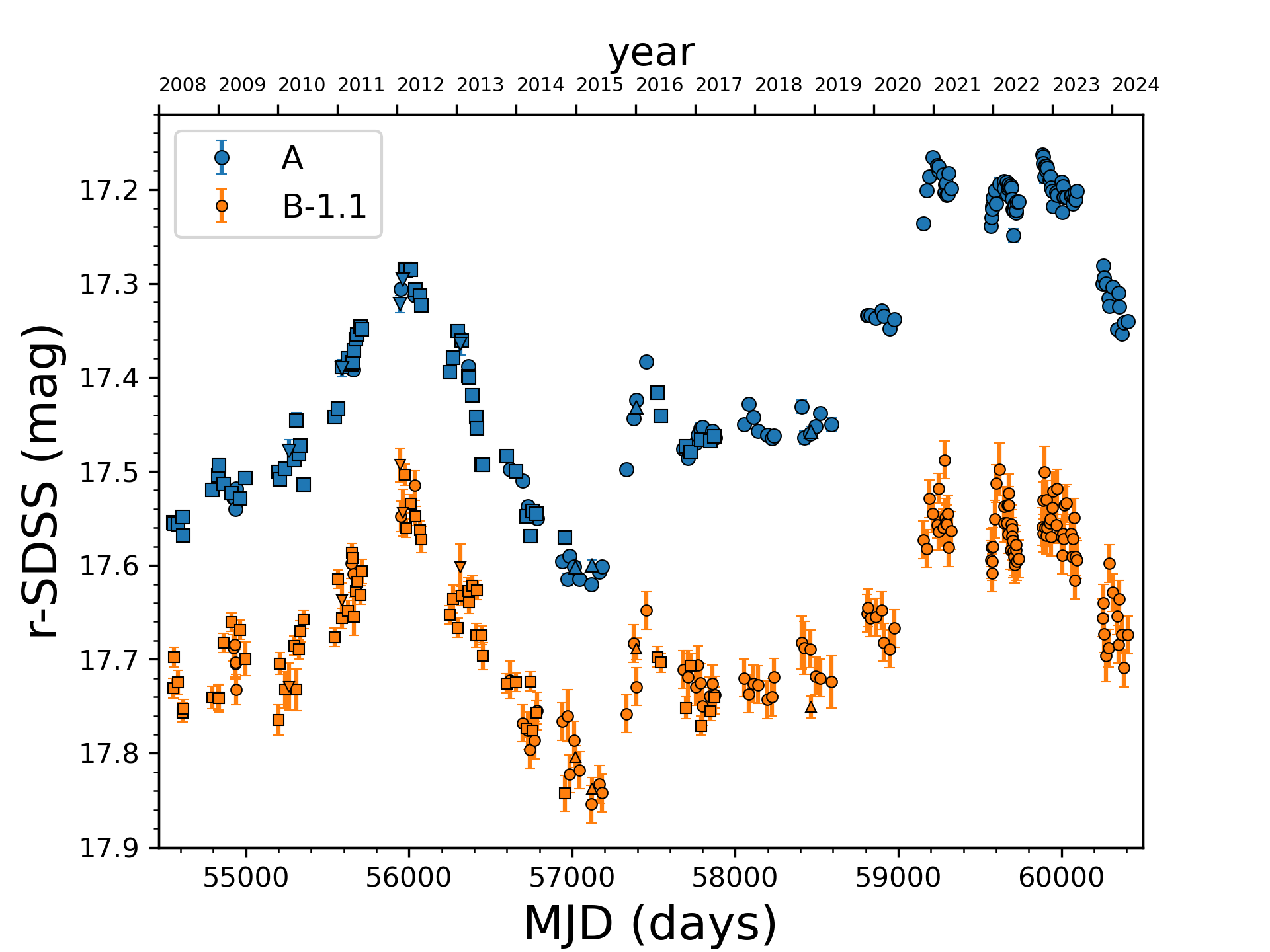}
\caption{GLENDAMA+ light curves of \object{FBQ 0951+2635}. These new variability records 
mainly consist of photometric data from the LT (circles) and KST (squares), but also 
incorporate information provided by the PS (inverted triangles) and DES (triangles) 
databases. See main text for details.}
\label{fig:newlcs}
\end{figure}

To improve the sampling over the period 2008$-$2024, we added several $r$-band magnitudes 
(LT photometric system) in the Data Release 1 of 
Pan-STARRS\footnote{http://panstarrs.stsci.edu} \citep[PS;][]{2020ApJS..251....7F} and the 
Data Release 2 of the Dark Energy Survey\footnote{https://www.darkenergysurvey.org} 
\citep[DES;][]{2021ApJS..255...20A}. The new 17-yr brightness records including magnitudes 
at 213 epochs are available in Table 1 at the CDS. These LT-KST-PS-DES (GLENDAMA+) light 
curves are also displayed in Figure~\ref{fig:newlcs}.

\subsection{Time delay measurement}
\label{sec:tdelmeas}

The $R$-band NOT light curves did not permit to measure a time delay between A and B with a 
precision of $\sim$10\%. The analysis of \citet{2005A&A...431..103J} suggested a delay 
interval of 10$-$20 d (depending on the technique, the smoothing, and the data points used), 
and subsequent studies indicated delay intervals of 10$-$30 d \citep{2011A&A...536A..44E} 
and even including negative values, i.e., image B leads image A \citep{2015A&A...580A..38R}.
Hence, additional light curves are required to address time delay measurements to 
$\sim$10\%, and here we consider the NOT dataset along with the GLENDAMA+ one. There is 
clear evidence for intrinsic variations in the 2008$-$2024 period, since brightness changes 
of A and B are significantly greater than photometric uncertainties, and additionally, 
A and B have an almost parallel behaviour (see Figure~\ref{fig:newlcs}). 

Because the optical light curves of \object{FBQ 0951+2635} are affected by a long-timescale 
microlensing episode \citep[e.g.,][]{2009MNRAS.397.1982S,2023ApJ...952...54R}, we measured 
the time delay using cross-correlation techniques that account for extrinsic variability. We 
have developed easy-to-use Python scripts to estimate the time delay of a doubly imaged 
quasar from two different methods incorporating polynomial microlensing variability. These 
scripts are publicly available at 
GitHub\footnote{https://github.com/glendama/q0951time\_delay} and briefly described in 
Appendix~\ref{sec:codes}. The first method we have used is the most sophisticated variant of 
the dispersion technique \citep[$D^2_4$;][]{1996A&A...305...97P}, which relied on a 
weighting scheme for the squared differences between the B image data and shifted A image 
data, where the A image data were shifted by a time lag $\tau$ and a microlensing polynomial 
of degree $N_{\rm{ml}}$. Our Gaussian weighting scheme was characterised by a parameter 
$\beta$ and is outlined in Appendix~\ref{sec:codes}. After setting values for $N_{\rm{ml}}$ 
and $\beta$, the dispersion spectrum $D^2_4$ was computed as the weighted sum of squared 
differences for time lags within a reasonable interval. To build this spectrum, we 
previously solved the microlensing polynomial coefficients that minimise it for each $\tau$ 
value. The time delay is assumed to be the time lag that minimises the dispersion spectrum. 

\begin{figure}
\centering
\includegraphics[width=9cm]{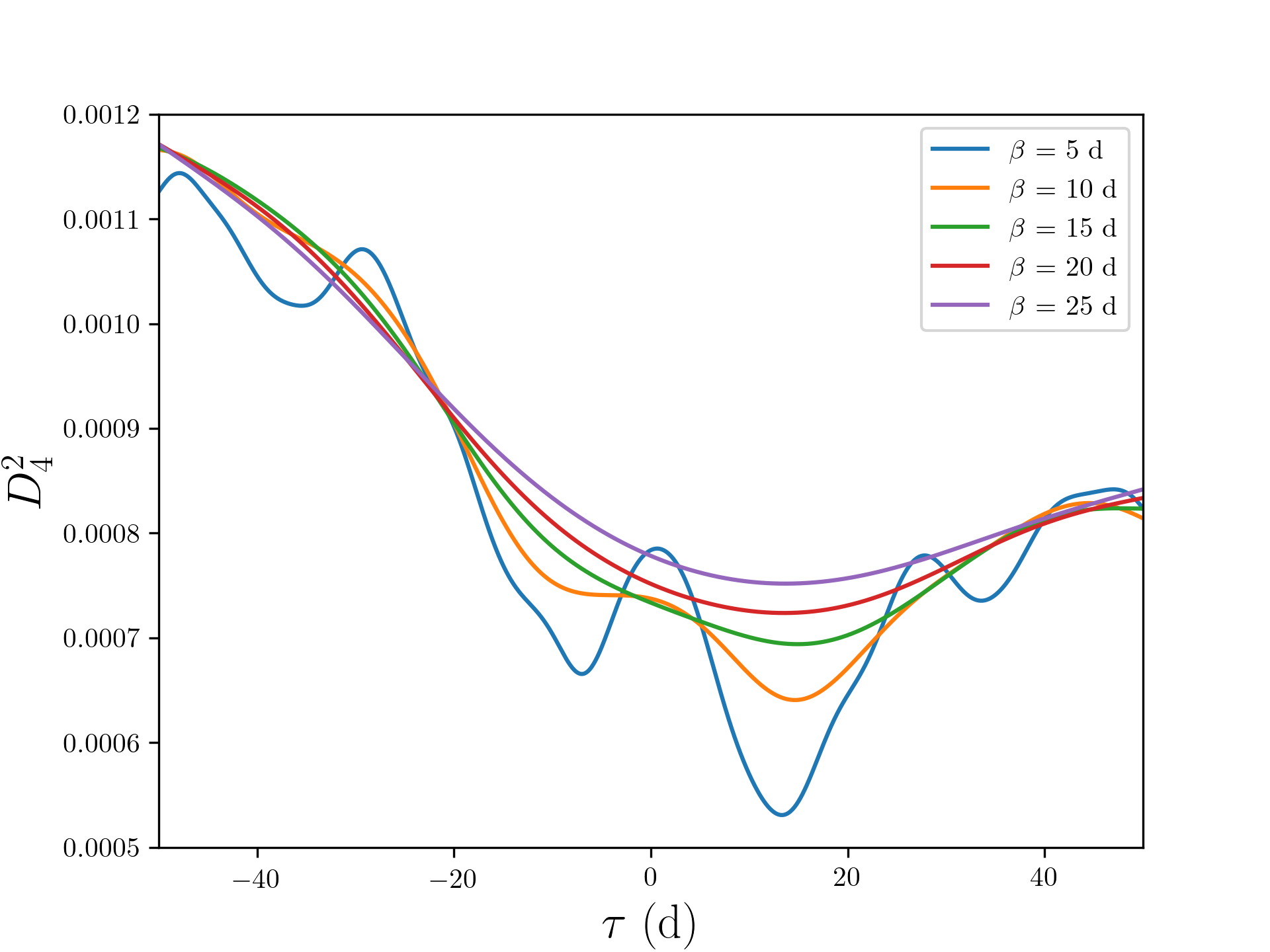}
\includegraphics[width=9cm]{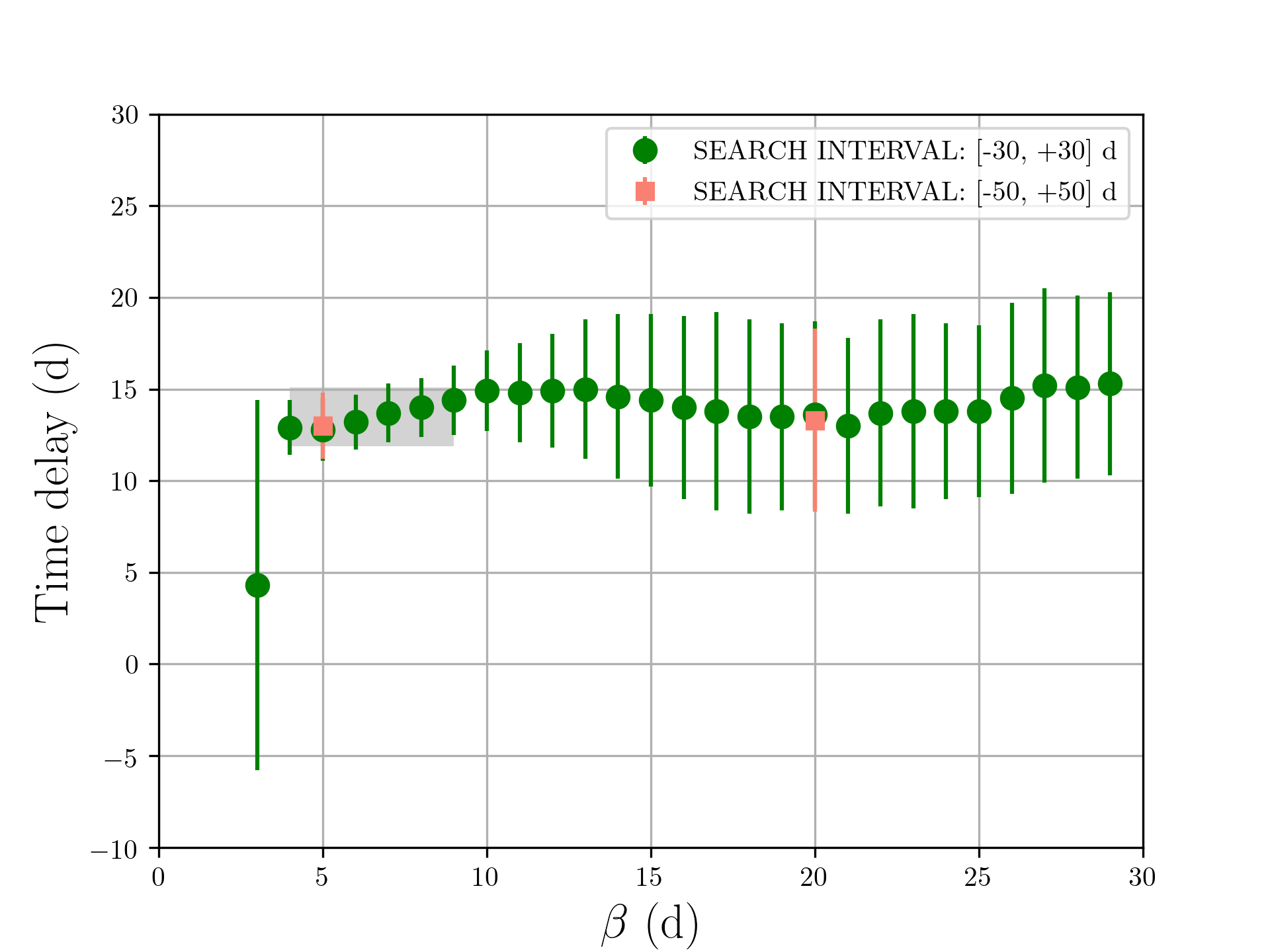}
\caption{Time delay estimation from the dispersion technique. Dispersion spectra $D^2_4$ are 
based on a Gaussian weighting function whose width is described by the parameter $\beta$, as
well as a microlensing polynomial of degree five ($N_{\rm{ml}}$ = 5; see main text). {\it 
Top panel}: $D^2_4$ spectra using the NOT light curves plus the new GLENDAMA+ brightness 
records and five values of $\beta$. These spectra have minima at time lags $\tau$ = 13$-$15 
d. {\it Bottom panel}: 1$\sigma$ confidence intervals for the time delay using 1000 
synthetic light curves of each image and a large set of $\beta$ values. A combined 
measurement from the six delay intervals for $\beta$ = 4$-$9 d is highlighted with a 
light-grey rectangle encompassing the individual measurements.}
\label{fig:disper}
\end{figure}

\begin{figure}
\centering
\includegraphics[width=9cm]{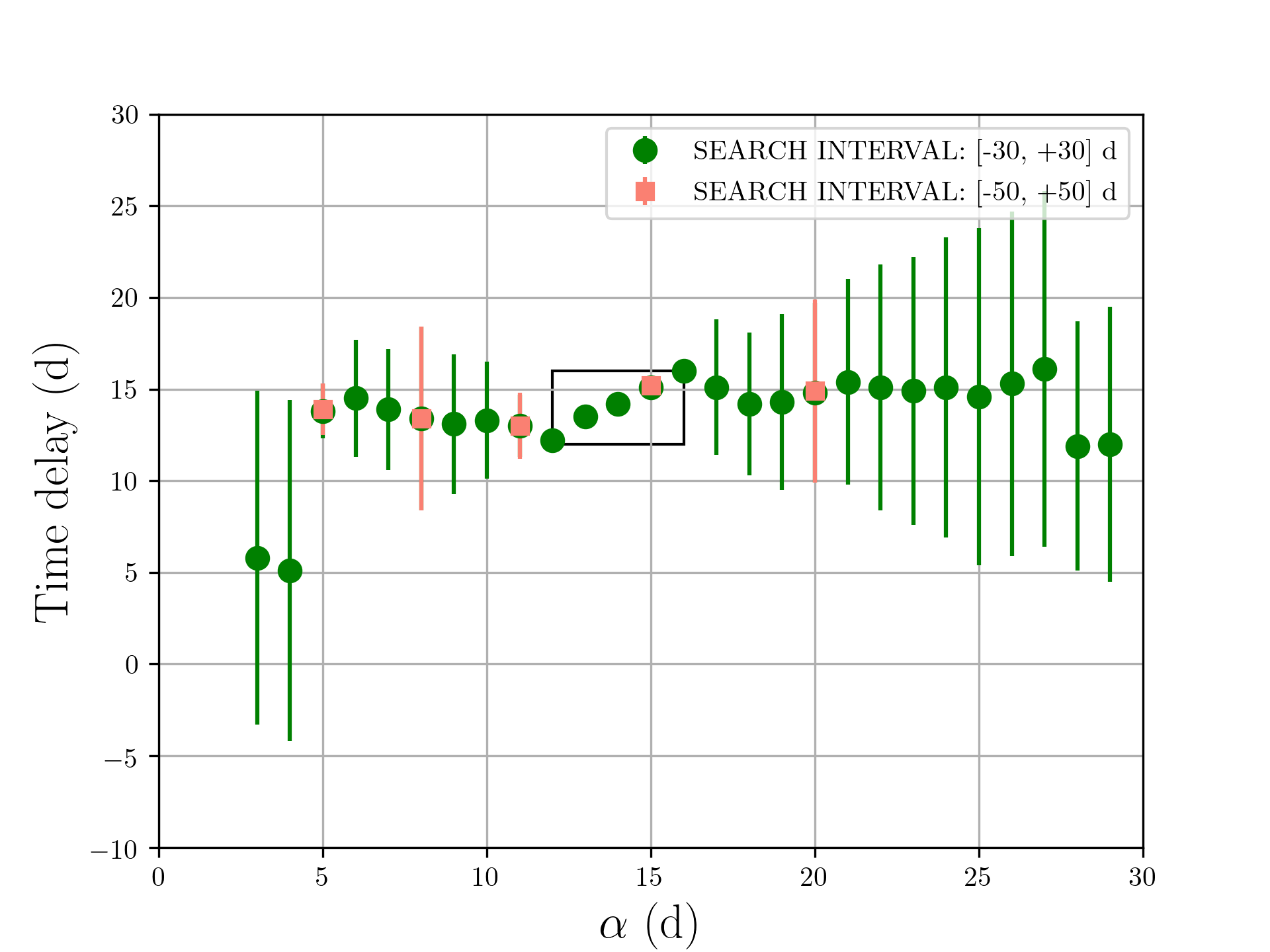}
\caption{Time delay estimation from the chi-square technique and 1000 synthetic light curves 
of each quasar image. To match light curves of the two images, a microlensing polynomial of 
degree five and binned light curves of B are considered, where bins of the B image have a 
semisize $\alpha$. The error bars represent 1$\sigma$ confidence intervals for the time 
delay, and the black outlined rectangle contains five anomalous measurements (see main 
text).}
\label{fig:chisqu}
\end{figure}

Considering polynomials of different degrees and several values of the parameter $\beta$, we 
verified that minimum dispersions decrease when the polynomial degree is progressively 
increased up to $N_{\rm{ml}}$ = 5, while a polynomial of degree $N_{\rm{ml}}$ = 6 does not 
produce substantial decreases in minimum dispersions. For $N_{\rm{ml}}$ = 5 and a broad 
range of $\beta$ values (3 $< \beta <$ 30 d), we found stable minima in the $D^2_4$ spectra 
covering the time lag interval [$-$50, +50] d. Some of these spectra with minima at 13$-$15 
d are shown in the top panel of Figure~\ref{fig:disper}. 

To quantify delay uncertainties, we generated 1000 synthetic light curves of each image 
based on the observed records \citep[e.g.,][]{2017ApJ...836...14S}. More precisely, we 
made 1000 "repetitions of the experiment" (synthetic curves for A and B) by adding random 
quantities to the measured magnitudes. These random quantities were realisations of normal 
distributions around zero, with standard deviations equal to the measured errors. We do not 
exactly repeat the original experiment but rather a worsened experiment with errors greater 
than those observed, that is, simulated magnitudes are not derived through a reconstruction 
of the underlying signal but from the observed one. Our simulation scheme is thus 
conservative regarding the size of error bars (about 20\% larger than the measured ones), 
with the caveat that only uncorrelated noise is added. We obtained distributions of time 
lags and polynomial coefficients that minimise the $D^2_4$ spectra from the simulated 
records. We then searched for minima within the interval [$-$30, +30] d to speed up 
numerical calculations involving thousands of dispersion spectra. However, for certain 
$\beta$ values, we checked that results using this narrower range of time lags are basically 
identical to those using the interval [$-$50, +50] d. 

For $\beta$ = 3 d, local minima in $D^2_4$ play a role, while for $\beta >$ 10 d, dispersion 
minima are not deep and delay uncertainties are relatively large. For $\beta$ = 4$-$9 d, the 
time delay values are around 13$-$14 d with 1$\sigma$ errors less than 2 d (see the bottom 
panel of Figure~\ref{fig:disper}). We have then combined (averaged) these six precise delay 
measurements to obtain the 1$\sigma$ confidence interval 13.5 $\pm$ 1.6 d (light-grey 
rectangle). The central values of the individual delays were averaged to find the mean 
value, and the uncertainty was computed as the square root of the sum of squares of two 
contributions: standard error of the mean value and average of individual delay errors.

The time delay of \object{FBQ 0951+2635} was also measured from a variant of the $\chi^2$ 
technique. The A image data were shifted by a time lag $\tau$ and a microlensing polynomial 
of degree $N_{\rm{ml}}$ = 5, and then compared with B image data binned around the 
time-shifted dates of A. The bins with semisize $\alpha$ were constructed through a linear 
weighting function, and for a given value of $\alpha$, we used a reduced chi-square 
($\chi^2_{\rm r}$) minimisation to estimate the polynomial coefficients and time lag that 
make the shifted light curve of A and the binned brightness record of B to match (see 
details in Appendix~\ref{sec:codes}). We remark that a polynomial of degree $N_{\rm{ml}} <$ 
5 leads to relatively high minimum $\chi^2_{\rm r}$ values, and the minimum $\chi^2_{\rm r}$ 
values for $N_{\rm{ml}}$ = 5 are barely modified by taking $N_{\rm{ml}}$ = 6. Thus, it is 
justified the choice $N_{\rm{ml}}$ = 5 when using the $\chi^2_{\rm r}$ method. The PyCS3 
software\footnote{https://gitlab.com/cosmograil/PyCS3} \citep[e.g.,][]{2013A&A...553A.120T,
2020A&A...640A.105M} includes another variant of the $\chi^2$ technique to estimate time 
delays between images of gravitationally lensed quasars. However, the intrinsic variability 
is modelled as a free-knot spline, so we adopted a simpler scheme avoiding to fit a complex 
intrinsic signal.

For 3 $< \alpha <$ 30 d, the 1000 simulated light curves of each quasar image (see above) 
allowed us to calculate 1$\sigma$ confidence intervals for the time delay (see 
Figure~\ref{fig:chisqu}). We searched again for minima within the interval [$-$30, +30] d, 
but checked the absence of border biases by using a second, more extended interval [$-$50, 
+50] d for several $\alpha$ values. Excessively small or large values of $\alpha$ (or 
$\beta$ for the minimum dispersion method; see above) are not suitable for time delay 
estimates. For $\alpha \geq$ 12 d, $\chi^2_{\rm r}$ minima are affected by systematics 
(sudden drops in $\chi^2_{\rm r}$ at certain time lags) or are not deep. The mentioned 
systematics are responsible for the anomalous delay-$\alpha$ relationship for $\alpha$ = 
12$-$16 d (see the black outlined rectangle in Figure~\ref{fig:chisqu}). This anomaly 
consists of a set of formally ultra-precise delays around 14 d that are not consistent with 
each other. For $\alpha$ = 5$-$11 d, delay estimates are consistent with results from the 
dispersion technique, and the two most precise measurements (for $\alpha$ = 5 and 11 d) are 
in excellent agreement with the dispersion-based delay of 13.5 $\pm$ 1.6 d. We also note 
that minima in $\chi^2_{\rm r}$ for $\alpha$ = 5 and 11 d are derived from $\sim$140 and 
$\sim$190 AB data pairs, respectively, so there is a significant overlap between records of 
A and B.

\subsection{Long-term microlensing}
\label{sec:longmicro} 

\begin{figure}
\centering
\includegraphics[width=9cm]{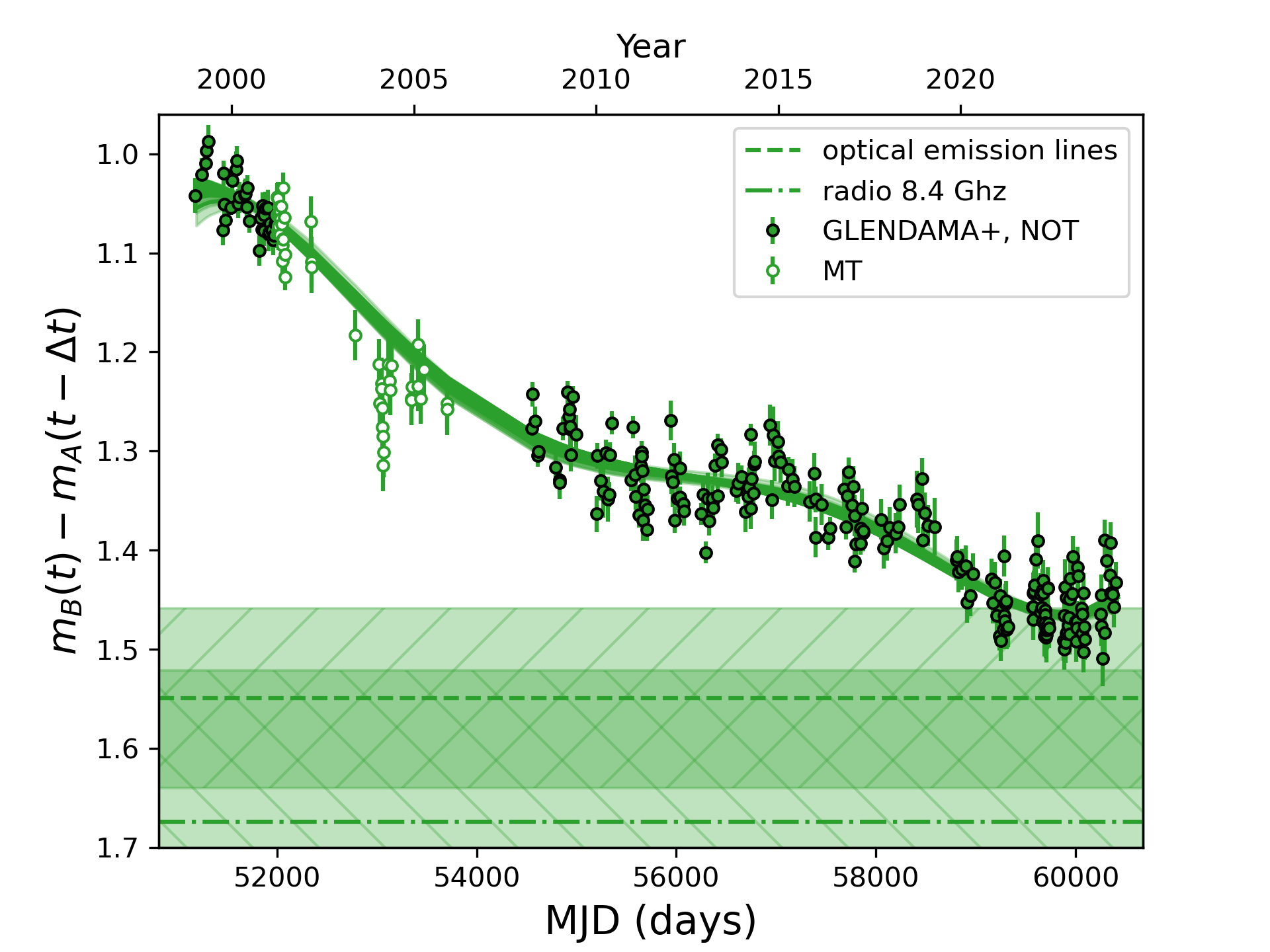}
\caption{New 25-yr DLC of \object{FBQ 0951+2635} in red optical passbands from NOT-MT and
GLENDAMA+ data. We also show solutions of the microlensing polynomial of degree five (thick 
solid line; see main text for details). For comparison purposes, we include magnitude 
differences from VLA data \citep[horizontal dash-dotted line;][]{2005A&A...431..103J} and 
emission lines (horizontal dashed line; see Sect.~\ref{sec:obscons}). The striped rectangles
represent uncertainties in these two differences.}
\label{fig:microl}
\end{figure}

To trace the long-timescale extrinsic (microlensing) signal in red optical passbands, we 
built the difference light curve (DLC) from 1999 to 2024 (see Figure~\ref{fig:microl}). This 
historical DLC relies on the $R$-band NOT-MT light curves and the GLENDAMA+ records in 
Figure~\ref{fig:newlcs}, and it consists of differences between the image B magnitudes at 
original epochs and image A magnitudes shifted by the time delay $\Delta t$ = 13.5 d, i.e., 
$m_{\rm{B}}(t) - m_{\rm{A}}(t - \Delta t)$. The differences were determined by linear 
interpolation of the time-shifted record of A. Figure~\ref{fig:microl} also displays 
dispersion-based solutions of the microlensing polynomial of degree five (thick solid line), 
which result from 1$\sigma$ confidence intervals of polynomial coefficients for 3 $< \beta 
<$ 30 d. Although these solutions were obtained using only NOT and GLENDAMA+ data (filled 
circles), they predict reasonably well the DLC shape at intermediate epochs \citep[open 
circles; however, as expected, the solutions do not account for the bump around day 53\,000 
reported by][]{2009MNRAS.397.1982S}. The DLC has evolved from an initial value close to 1 
mag at the end of the last century to 1.4$-$1.5 mag in the current decade. Hence, some 
recent DLC values are roughly consistent with the $B/A$ flux ratio from radio observations 
and emission lines (see Sect.~\ref{sec:obscons}).  

\section{Observational constraints and structure of the main lensing galaxy}
\label{sec:struc}

\subsection{Observational constraints}
\label{sec:obscons}

\setcounter{table}{1}
\begin{table}[h!]
\begin{center}
\caption{Summary of observational constraints of \object{FBQ 0951+2635}.}
\label{tab:constra}
\begin{tabular}{lcc}
   \hline \hline
   Parameter (units) & Constraint & Reference\\
   \hline 
   $z_{\rm{s}}$ 		       & 1.249                  & 1\\
   $z_{\rm{l}}$  		       & 0.260                  & 2\\
   	\multicolumn{1}{r}{APS 1}      &		        &  \\
   $x_{\rm A}$ (\arcsec)               & 0.000 $\pm$ 0.001      & 3\\
   $y_{\rm A}$ (\arcsec)               & 0.000 $\pm$ 0.001      & 3\\                     
   $x_{\rm B}$ (\arcsec)               & $-$0.892 $\pm$ 0.001   & 3\\
   $y_{\rm B}$ (\arcsec)               & $-$0.628 $\pm$ 0.001   & 3\\
   $x_{\rm G}$ (\arcsec)               & $-$0.750 $\pm$ 0.002   & 3\\
   $y_{\rm G}$ (\arcsec)               & $-$0.459 $\pm$ 0.002   & 3\\	  
   $e_{\rm{G}}$  		       &  0.25 $\pm$ 0.04       & 3\\
   $\theta_{e_{\rm G}}$ (\degr)	       &  22 $\pm$ 4            & 3\\
   $R_{\rm{G}}$ (\arcsec)              &  0.09 $\pm$ 0.02       & 3\\
	\multicolumn{1}{r}{APS 2}      &		        &  \\
   $x_{\rm A}$ (\arcsec)               & 0.0000 $\pm$ 0.0001    & 4\\
   $y_{\rm A}$ (\arcsec)               & 0.0000 $\pm$ 0.0001    & 4\\
   $x_{\rm B}$ (\arcsec)               & $-$0.8983 $\pm$ 0.0012 & 4\\
   $y_{\rm B}$ (\arcsec)               & $-$0.6336 $\pm$ 0.0012 & 4\\
   $x_{\rm G}$ (\arcsec)               & $-$0.7521 $\pm$ 0.0028 & 4\\
   $y_{\rm G}$ (\arcsec)               & $-$0.4603 $\pm$ 0.0028 & 4\\	  
   $e_{\rm{G}}$  		       &  0.47 $\pm$ 0.03       & 4\\
   $\theta_{e_{\rm G}}$ (\degr)	       &  12.8 $\pm$ 2.2        & 4\\
   $R_{\rm{G}}$ (\arcsec)              &  0.78 $\pm$ 0.01       & 4\\	
				       &                        &  \\
   $M$    		 	       &  0.23 $\pm$ 0.02       & 5\\
   $\Delta t$ (d)                      &  13.5 $\pm$ 1.6        & This paper\\
   $\kappa_{\rm ext}$                  &  0.23                  & 6\\
   $\gamma_{\rm ext}$                  &  0.04                  & 6\\
   $\theta_{\gamma_{\rm ext}}$ (\degr) &  $-$44                 & 6\\
   \hline
\end{tabular}
\end{center}
\footnotesize{Note: Here, $z_{\rm{s}}$ and $z_{\rm{l}}$ are the redshifts of the double 
quasar (images A and B) and the main lensing galaxy G, respectively. Additionally, regarding 
the two astro-photometric solutions APS 1 and APS 2, the brightest image A is located at the 
origin of coordinates, and the positive direction of $x$ is defined by west, while the 
positive direction of $y$ is defined by north. Some astrometric errors have also been made 
symmetric ($\sigma_x$ = $\sigma_y$). The morphology of G (De Vaucouleurs brightness profile) 
is given by three parameters: ellipticity ($e_{\rm{G}}$), position angle of the major axis 
($\theta_{e_{\rm G}}$; it is measured east of north), and effective radius ($R_{\rm{G}}$). 
We also include the macro-magnification ratio $M$ and time delay $\Delta t$ with respect to 
the image A. All astro-photometric, magnification and delay constraints are 1$\sigma$ 
confidence intervals. The last three rows correspond to the external convergence, external 
shear strength, and position angle of the external shear. We also assume a flat $\Lambda$CDM 
cosmology (see main text).}
\tablebib{
(1) \citet{2015ApJS..219...29M}; 
(2) \citet{2007A&A...465...51E}; 
(3) \citet{2005A&A...431..103J}; 
(4) \citet{2012A&A...538A..99S};
(5) \citet{2012A&A...544A..62S}; 
(6) \citet{2017ApJ...850...94W}.
}
\end{table}

We focus on the double quasar \object{FBQ 0951+2635}, giving here details on its 
observational constraints and summarizing them in Table~\ref{tab:constra}. The spectroscopic 
redshifts of the source and the main lensing galaxy G are cited in Sect.~\ref{sec:introd}.
\citet{2005A&A...431..103J} and \citet{2012A&A...538A..99S} also used near-IR HST imaging 
\citep{2000ApJ...543..131K} to determine the relative astrometry of the system and the 
morphology of the early-type galaxy G, whose brightness profile was modelled by a De 
Vaucouleurs law. However, the two independent analyses of the same data yielded two 
astro-photometric solutions (APS 1 and APS 2) that differ in image separation, i.e., angular 
distance between A and B, and galaxy morphological parameters (see Table~\ref{tab:constra}). 
Additionally, the discovery paper \citep{1998AJ....115.1371S} reported VLA radio fluxes of 
both quasar images, which led to a radio flux ratio $B/A$ = 0.21 $\pm$ 0.03  
($m_{\rm{B}} - m_{\rm{A}}$ = 1.69 $\pm$ 0.16 mag) at 8.4 GHz \citep{2005A&A...431..103J}. 

Using optical spectra of \object{FBQ 0951+2635} from the Keck II Telescope 
\citep{1998AJ....115.1371S}, the Very Large Telescope \citep{2007A&A...465...51E}, and the 
William Herschel Telescope \citep{2021A&A...653A.109F}, we also estimated the flux ratio 
$B/A$ for the cores of the C\,{\sc iv}, C\,{\sc iii}], and Mg\,{\sc ii} emission lines. The 
six $B/A$ values range bewteen 0.20 and 0.26, and a basic statistics of these six estimates 
(mean = 0.24 and standard deviation = 0.02, or equivalently, $m_{\rm{B}} - m_{\rm{A}}$ 
= 1.55 $\pm$ 0.09 mag) is in good agreement with the radio flux ratio (see above) and the 
macro-magnification ratio $M$ in Table 4 of \citet{2012A&A...544A..62S}. We then adopted 
Sluse et al.'s constraint on $M$. In addition to the spectroscopic test of the 
macro-magnification ratio, this paper presents a time delay measurement to $\sim$10\% by 
combining early optical light curves and those through observations made over the last 17 yr
(see Sect.~\ref{sec:tdelmeas}). In a near future, better photometric data in more densely 
sampled light curves might lead to a delay error less than 1.5 d. However, taking a possible
microlensing-induced scatter of $\sim$1 d into account \citep[e.g.,][]{2018MNRAS.473...80T}, 
this perspective may be excessively optimistic.

The time delay between the two images of a double quasar is basically given by the product 
of the time-delay distance and the Fermat potential variation between both images 
\citep[e.g.,][]{2016A&ARv..24...11T,2018SSRv..214...91S}. The Fermat potential variation 
relies upon the lensing mass distribution. In addition, the time-delay distance depends on 
$z_{\rm{s}}$ and $z_{\rm{l}}$, and it is inversely proportional to $H_0$, with other 
cosmological parameters playing a lesser role. For consistency with current constraints on
secondary deflectors (see here below), we adopted a flat $\Lambda$CDM cosmology with $H_0$ = 
71 km s$^{-1}$ Mpc$^{-1}$, $\Omega_{\rm{M}}$ = 0.274, and $\Omega_{\Lambda}$ = 0.726
\citep{2009ApJS..180..225H,2009ApJS..180..330K}. Using these cosmological parameters, 
\citet{2017ApJ...850...94W} focused on determining the gravitational lensing effect of 
galaxy groups for \object{FBQ 0951+2635}, i.e., the group to which G belongs (environment of 
G) and additional line-of-sight groups. They did not explore the minor influence of isolated 
galaxies on the lensing potential, and estimated the typical values of the external 
convergence ($\kappa_{\rm ext}$), external shear strength ($\gamma_{\rm ext}$), and position 
angle of the external shear ($\theta_{\gamma_{\rm ext}}$) that are shown in 
Table~\ref{tab:constra}. 

\subsection{Methodology to reconstruct the mass distribution of the main deflector}
\label{sec:methodmass}

The lensing mass distribution was initially modelled using two standard components: a 
singular power-law ellipsoid (SPLE) to describe the main lensing galaxy G and an external 
shear caused by the additional mass intervening \citep[e.g.,][]{2013ApJ...766...70S,
2019MNRAS.483.5649S}. Although a SPLE is often characterised by its logarithmic mass-density 
slope $\gamma$, in this paper, we used the power-law index $\alpha_{\rm pl}$ ($\alpha_{\rm 
pl} = 3 - \gamma$) to avoid any confusion with the shear and to be consistent with the 
notation in the model catalog of the GRAVLENS/LENSMODEL software \citep{2001astro.ph..2340K,
2001astro.ph..2341K}. Despite neither the light nor the dark matter distribution in galaxies 
follow a power-law radial profile, the combined (total) density profile of G is expected to 
be close to isothermal \citep[SPLE with $\alpha_{\rm pl} \sim$ 1; 
e.g.,][]{2006ApJ...649..599K,2015ApJ...804L..21C,2020MNRAS.491.5188W,2024arXiv240810316S}. 
It is also noteworthy that the quasi-isothermal combined model leads to similar results than 
a model treating light and dark matter individually \citep[e.g.,][]{2014ApJ...788L..35S,
2020A&A...639A.101M}. For comparison purposes, we also considered a constant mass-to-light 
ratio model of G, i.e., a De Vaucouleurs (DV) surface mass density instead of that for a 
SPLE.

In addition to the power-law index, the parameters of the SPLE model are the 2D position 
($x_{\rm 0}$, $y_{\rm 0}$), mass scale\footnote{The mass scale is defined in Eq. (26) 
of \citet{2001astro.ph..2341K}. For a singular isothermal sphere, it equals the Einstein 
radius} $b$, ellipticity $e$, and position angle $\theta_e$. The external shear is also 
described by two parameters: strength $\gamma_{\rm ext}$ and position angle 
$\theta_{\gamma_{\rm ext}}$, whereas it is not necessary to explicitly incorporate the 
external convergence into the lensing mass model. However, if $\kappa_{\rm ext} \neq 0$, 
some model parameters must be appropriately reinterpreted 
\citep[e.g.,][]{1985ApJ...289L...1F,1996ApJ...464...92G}. We mean $b \mapsto b^{\ast} = b/(1 
- \kappa_{\rm ext})^{1/(2 - \alpha_{\rm pl})}$ and $\gamma_{\rm ext} \mapsto \gamma_{\rm 
ext}^{\ast} = \gamma_{\rm ext}/(1 - \kappa_{\rm ext})$, as well as $H_0 \mapsto H_0^{\ast} = 
H_0/(1 - \kappa_{\rm ext})$ \citep[e.g.,][]{2010ApJ...711..246F}. For example, for this 
system with such high external convergence (see Table~\ref{tab:constra}), the use of 
$H_0^{\ast}$ (instead of $H_0$) plays a critical role when the time delay is considered as a 
constraint on the mass distribution.

\begin{figure*}
\centering
\includegraphics[width=9cm]{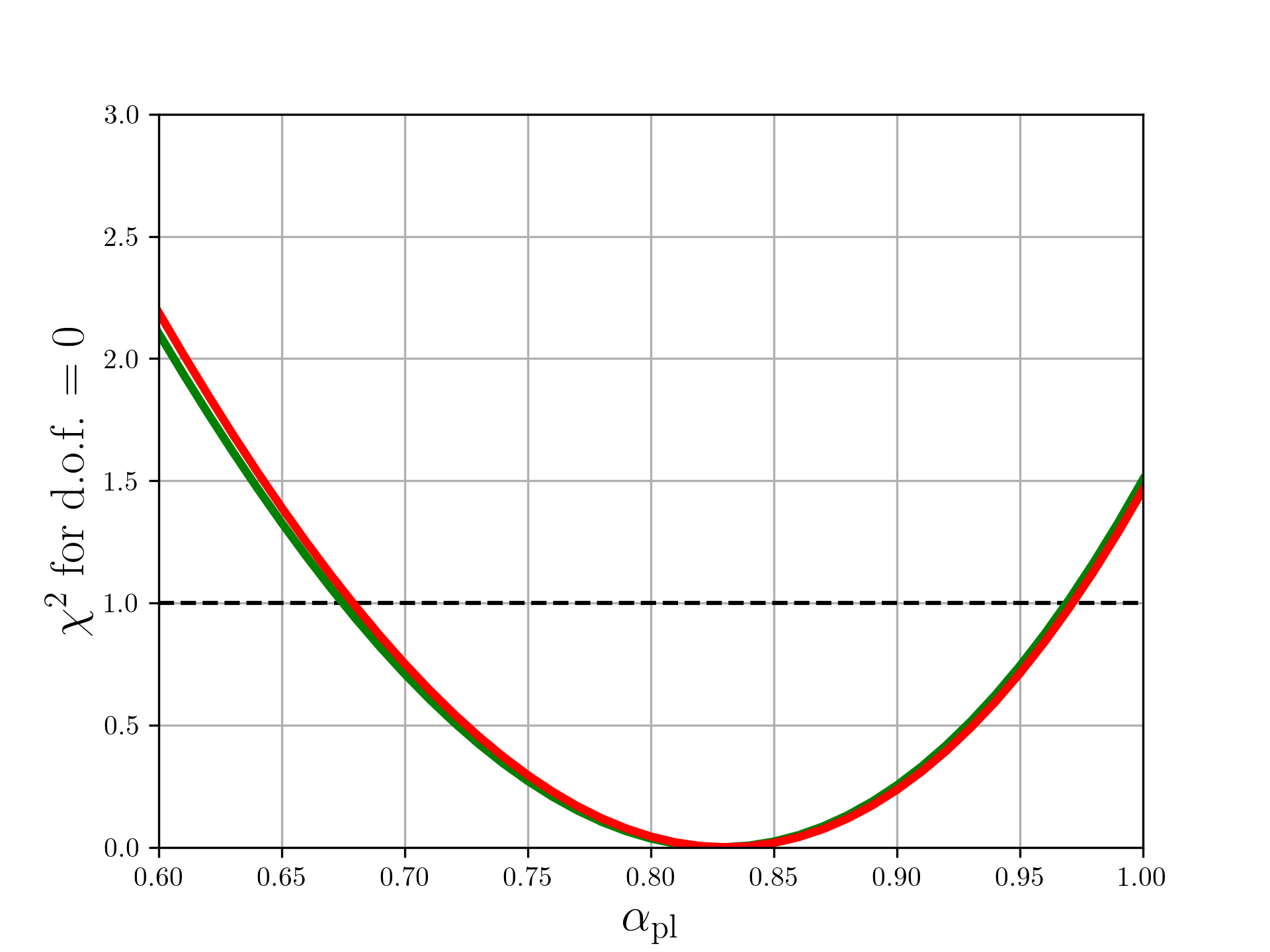}
\includegraphics[width=9cm]{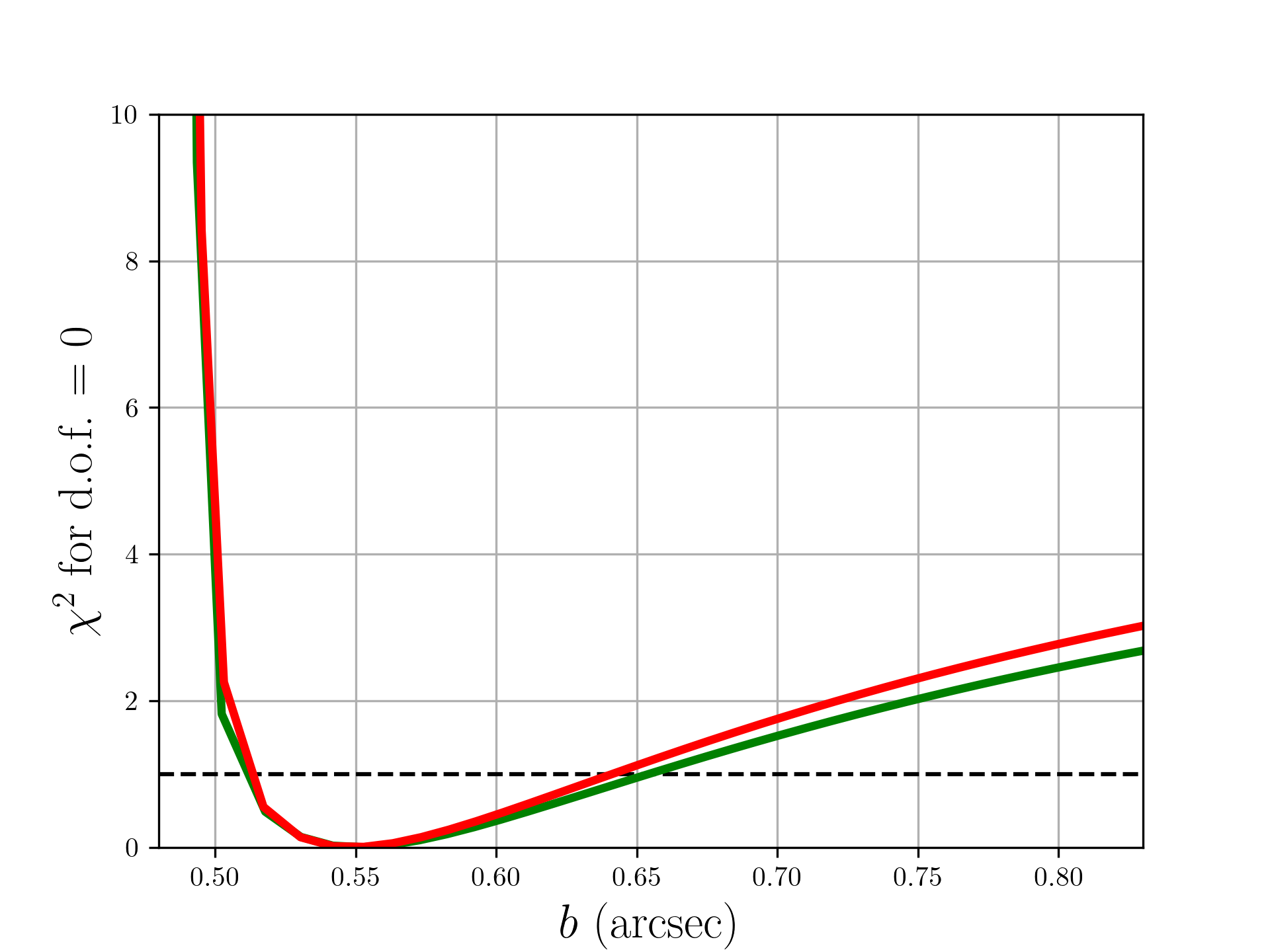}
\includegraphics[width=9cm]{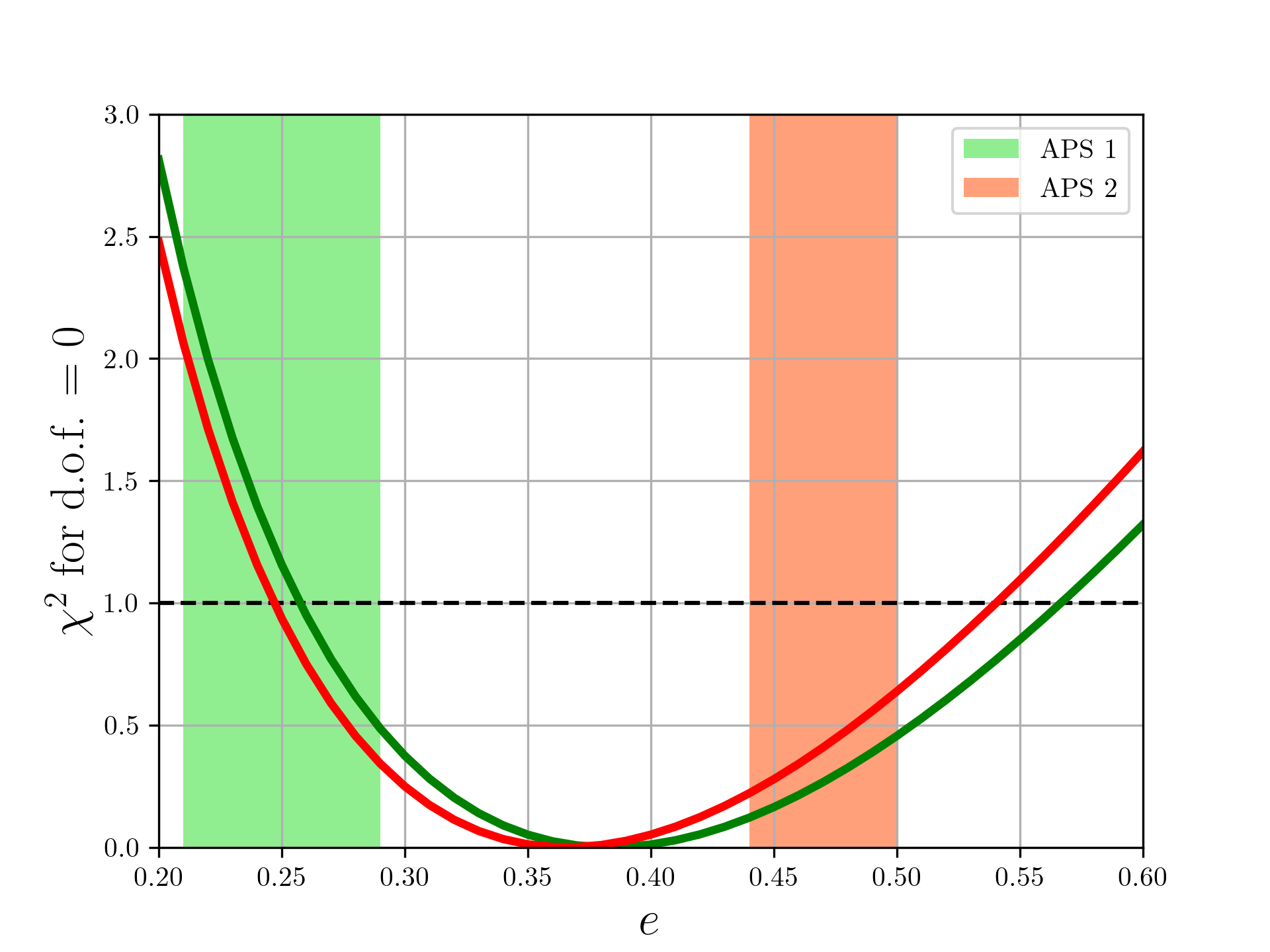}
\includegraphics[width=9cm]{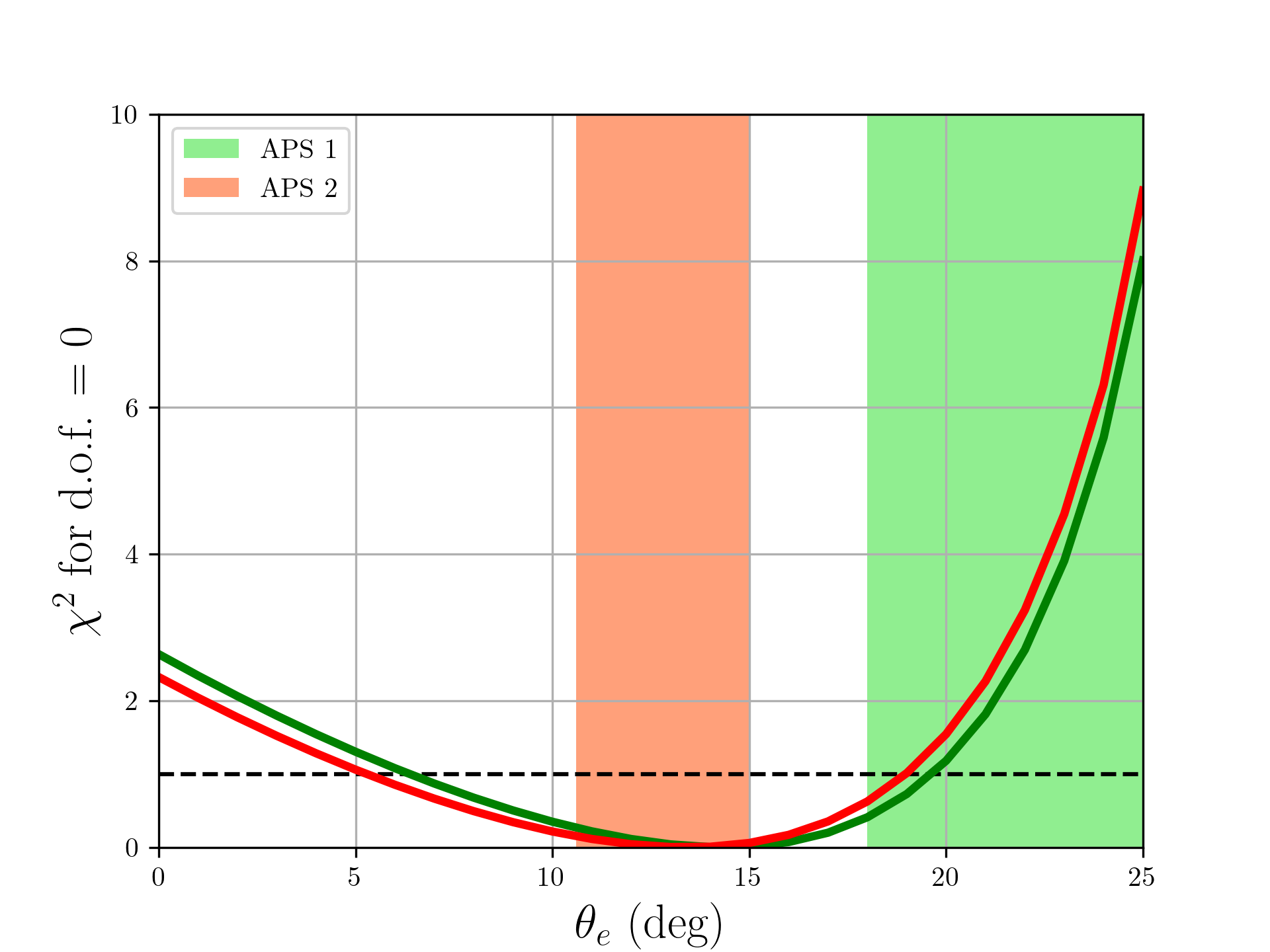}
\caption{SPLE parameters for the galaxy acting as main gravitational deflector to 
\object{FBQ 0951+2635}. Lines in green and red result from the astrometric constraints in 
APS 1 \citep[see Table~\ref{tab:constra};][]{2005A&A...431..103J} and APS 2 \citep[see 
Table~\ref{tab:constra};][]{2012A&A...538A..99S}, respectively. {\it Top left panel}: 
power-law index. {\it Top right panel}: true mass scale. {\it Bottom left panel}: 
ellipticity of the mass distribution. {\it Bottom right panel}: position angle of the mass 
distribution. Our solutions for $e$ and $\theta_e$ are also compared with the observed 
(near-IR) morphology of the galaxy. The light green and light salmon rectangles represent 
1$\sigma$ intervals in APS 1 and APS 2, respectively.}
\label{fig:splres}
\end{figure*}  

Setting the values of $z_{\rm{s}}$, $z_{\rm{l}}$, $\gamma_{\rm ext}^{\ast}$, 
$\theta_{\gamma_{\rm ext}}$, $H_0^{\ast}$, $\Omega_{\rm{M}}$, and $\Omega_{\Lambda}$, and 
taking 1$\sigma$ confidence intervals for the positions of A, B and G, the 
macro-magnification, and the time delay (see Sect.~\ref{sec:obscons}), the 
GRAVLENS/LENSMODEL software allowed us to fit the six parameters of the SPLE model (see 
above). Because we considered two astro-photometric solutions (APS 1 and APS 2), we obtained 
two different reconstructions of the SPLE mass distribution. We note again that it is the 
effective mass scale $b^{\ast}$ which we fit in our model, so the true mass scale is derived 
from the product of $b^{\ast}$, and a scale factor depending on the parameters $\kappa_{\rm 
ext}$ and $\alpha_{\rm pl}$. In the least-squares fitting of the SPLE, the number of degrees 
of freedom (d.o.f.) was zero. Additionally, the constant mass-to-light ratio model (DV 
model) of G also contains six parameters: $x_{\rm 0}$, $y_{\rm 0}$, $b^{\ast}$, $e$, 
$\theta_e$, and the effective radius $R$. However, in this scenario, since light traces 
mass, we constrained the morphology of the DV mass ellipsoid from the 1$\sigma$ confidence 
intervals of $e_{\rm{G}}$, $\theta_{e_{\rm G}}$, and $R_{\rm{G}}$ in 
Table~\ref{tab:constra}. Thus, as a result of adding three more constraints, d.o.f. = 3.

\subsection{Results}
\label{sec:results}

Using the SPLE mass model and astrometric constraints in APS 1, the 1$\sigma$ intervals of 
the power-law index, true mass scale, ellipticity and position angle are $\alpha_{\rm pl}$ = 
0.822 $\pm$ 0.147, $b$ (\arcsec) = 0.585 $\pm$ 0.073, $e$ = 0.412 $\pm$ 0.155, and 
$\theta_e$ (\degr) = 13.0 $\pm$ 6.6. These results are very similar to those using 
astrometric constraints in APS 2: $\alpha_{\rm pl}$ = 0.825 $\pm$ 0.146, $b$ (\arcsec) = 
0.577 $\pm$ 0.064, $e$ = 0.394 $\pm$ 0.147, and $\theta_e$ (\degr) = 12.2 $\pm$ 6.9. In 
both cases, we obtain best fits with $\chi^2 \sim$ 10$^{-5}$ (d.o.f. = 0). 
Figure~\ref{fig:splres} also depicts $\chi^2$$-$$\alpha_{\rm pl}$, $\chi^2$$-$$b$, 
$\chi^2$$-$$e$, and $\chi^2$$-$$\theta_e$ relationships from the two astrometric datasets. 
As expected, there is only a small deviation from isothermality ($\alpha_{\rm pl}$ = 1; see 
Sect.~\ref{sec:methodmass}). Additionally, light and mass are closely aligned, since the 
mass ellipticity and its orientation agree reasonably well with their near-IR counterparts. 

We have shown that the SPLE-based reconstruction of the galaxy mass is very weakly 
influenced by the difference between image separation in APS 1 and APS 2. Moreover, despite 
discrepancies between the galaxy shape in these two datasets ($\sim$6$\sigma$ in 
$e_{\rm{G}}$ and $\sim$3$\sigma$ in $\theta_{e_{\rm G}}$), if formal errors are taking into 
account, both ($e_{\rm{G}}$, $\theta_{e_{\rm G}}$) measurements are consistent with the 
($e$, $\theta_e$) solutions from the SPLE mass model (see Figure~\ref{fig:splres}). Our 
lensing mass modelling focuses on the use of a Hubble constant that corresponds to an 
average of the main measurements from different experiments \citep[$H_0$ = 71 km s$^{-1}$ 
Mpc$^{-1}$; e.g.,][]{2021CQGra..38o3001D} and typical values of the constraints on secondary 
deflectors for this concordance $H_0$. Therefore, accounting for current uncertainties in 
$H_0$ and the effects by secondary deflectors increases the error bars of $\alpha_{\rm pl}$, 
$b$, $e$, and $\theta_e$, but does not modify their best-fit values. As an example, a 
reasonable $H_0$ interval of 68 to 74 km s$^{-1}$ Mpc$^{-1}$ 
\citep[e.g.,][]{2021CQGra..38o3001D} increases uncertainties in $\alpha_{\rm pl}$ by about 
6.5\%. These enlargements are similar to those associated with the errors in $\kappa_{\rm 
ext}$ and $\gamma_{\rm ext}$ \citep[the error in $\theta_{\gamma_{\rm ext}}$ is not 
mentioned;][]{2017ApJ...850...94W}, and clearly below the 18\% uncertainties estimated for 
$\alpha_{\rm pl}$.   

We also considered the DV mass model and all (astrometric and morphological) 
constraints in APS 1 and APS 2. Interestingly, the DV mass distribution is reasonably 
consistent with the constraints in APS 2 ($\chi^2 \sim$ 3.7), suggesting that a constant 
mass-to-light ratio galaxy cannot be ruled out \citep[see also][]{2013arXiv1303.6896M}. The 
best-fit solution includes a true mass scale of 1\farcs12 and reproduces well the 
morphological measurements of Sluse et al. ($e$ = 0.45, $\theta_e$ = 13.0\degr, and $R$ =
0\farcs78). However, using APS 1, we do not achieve a good solution for the lensing mass 
because $\chi^2 \sim$ 175 is much higher than d.o.f. = 3 (see the end of 
Sect.~\ref{sec:methodmass}). This high $\chi^2$ value is largely due to the morphological 
constraints in APS 1, and the main difference between the two light distributions of G lies 
in the effective radius: $R_{\rm{G}}$ in APS 2 is one order of magnitude larger than 
$R_{\rm{G}}$ in APS 1.

To decide whether or not light is an acceptable tracer of the mass in the galaxy at 
redshift 0.26, we have reanalysed the HST archival data of \object{FBQ 0951+2635}, paying 
special attention to the value of the effective radius. The central idea was to repeat 
Jakobsson et al.'s evaluation of $R_{\rm{G}}$, who used a technique less refined than that 
of the COSMOGRAIL project \citep{2012A&A...538A..99S}. More specifically, we taken the 
NICMOS drizzled frame in the $H$ band and used the A image as an empirical point spread 
function (as Jakobsson et al.'s did), and then applied the IMFITFITS software 
\citep{1998AJ....115.1377M} by setting the positions of A and B at the values cited by 
\citet{2005A&A...431..103J}. For a DV brightness profile, our approach yields an effective
radius significantly larger than that estimated by Jakobsson et al. and closer to Sluse at 
al.'s measurement. Jakobsson et al. most likely underestimated the effective radius, while 
Sluse et al.'s solution for $R_{\rm{G}}$ relies on an iterative MCS deconvolution algorithm 
\citep{2007A&A...470..467C} and seems much more reliable. Hence, we cannot rule out a 
constant mass-to-light ratio DV model.

\section{Conclusions}
\label{sec:concl}

Prior to this work, optical, near-IR and radio observations towards the doubly imaged quasar 
\object{FBQ 0951+2635} have provided a rich set of constraints for the gravitational lens 
system. However, unfortunately, early optical monitoring in the period 1999$-$2001 did not 
lead to a reliable time delay between the two quasar images 
\citep[e.g.,][]{2005A&A...431..103J,2011A&A...536A..44E,2015A&A...580A..38R}, which is a key 
constraint to "blindly" reconstruct the mass structure of the main lensing galaxy G, i.e., 
its mass scale and morphology, or to be used for time-delay cosmography  
\citep[e.g.,][]{2022arXiv221010833B}. Here, we focus on a robust determination of the time 
delay and a discussion of the structure of the non-local early-type galaxy G
\citep[e.g.,][]{2000ApJ...543..131K,2007A&A...465...51E}. 

We consider new optical light curves in the period 2008$-$2024, including significant 
intrinsic fluctuations and consisting of quasar image magnitudes at 213 epochs. These 
GLENDAMA+ records are merged together with the early ones, and we then use two different 
cross-correlation techniques to match the light curves of both images by allowing for a time
delay and microlensing polynomial variability. We also present easy-to-use Python scripts
associated with the two techniques and obtain a time delay of 13.5 $\pm$ 1.6 d (1$\sigma$ 
confidence interval). This reliable measurement to $\sim$10\% resolves the long-standing 
controversy over the time delay of \object{FBQ 0951+2635}. In addition, a microlensing 
polynomial of degree five describes well the long-timescale evolution of the difference 
light curve. We also note that the current optical-passband flux ratio is close to the 
macro-magnification ratio. 

Most previous reconstructions of the mass of G from a SPLE model were based on the 
assumption of isothermality ($\alpha_{\rm pl}$ = 1), and priors on $e$ and $\theta_e$ from 
the observed light structure \citep[e.g.,][]{2012A&A...538A..99S}. However, it is possible 
to "blindly" fit parameters of the mass structure of G by knowing 1$\sigma$ intervals for 
the positions of quasar images and G, the macro-magnification ratio, and the time delay, 
along with typical values of the source and lens redshifts, the convergence and shear due to 
secondary gravitational deflectors, and the main cosmological parameters. We mean that for a 
SPLE mass model, hypotheses/priors on $\alpha_{\rm pl}$, $e$ and $\theta_e$ are not 
required. Additionally, the effective mass scale arising from the SPLE model fitting can be 
converted into a true mass scale of G. Thus, the constraints in Table~\ref{tab:constra} lead 
to SPLE mass solutions with a deviation from isothermality consistent with studies of 
lensing galaxies selected from the SLACS Survey \citep{2006ApJ...649..599K,
2009MNRAS.399...21B,2009ApJ...703L..51K}. As expected, we also find good alignment between 
the mass distribution and the near-IR light of G \citep[e.g.,][]{2006ApJ...649..599K}. 
We also note that a DV mass model (constant mass-to-light ratio galaxy) cannot be ruled out 
from current data.  

The observational constraints in Table~\ref{tab:constra} can be used to take a deeper look 
at the mass structure of G by considering models of the galaxy other than the SPLE and 
DV \citep[e.g.][]{2013A&A...559A..37S,2020MNRAS.493.1725K}. For example, light-based priors 
on $e$ and $\theta_e$ allow one to study mass models incorporating two more parameters. New
observations of the lens system will also contribute to a better understanding of the galaxy 
structure, opening the door to the analysis of its mass distribution from accurate 
non-parametric reconstructions \citep[e.g.,][]{1997MNRAS.292..148S,2005MNRAS.360..477D}. 
Future observations should include spatially resolved stellar kinematics of G, and detailed 
imaging of the lensing galaxy and the lensed host galaxy of the quasar. In particular, 
new near-IR imaging along with state-of-the-art analysis techniques should definitively 
resolve the controversy over the light structure of the galaxy \citep{2005A&A...431..103J,
2012A&A...538A..99S,2023ApJ...952...54R}.

\begin{acknowledgements}
The Liverpool Telescope (LT) is operated on the island of La Palma by Liverpool John Moores 
University in the Spanish Observatorio del Roque de los Muchachos of the Instituto de 
Astrof\'isica de Canarias with financial support from the UK Science and Technology 
Facilities Council. We thank the staff of the LT for a kind interaction. We also thank the 
anonymous referee for providing valuable feedback on the original manuscript. VNS acknowledges 
the Universidad de Cantabria (UC) and the Spanish Agencia Estatal de Investigacion (AEI) for
financial support for a long stay at the UC in the period 2022–2024. This research has been 
supported by the grant PID2020-118990GB-I00 funded by MCIN/AEI/10.13039/501100011033. 
\end{acknowledgements}

\begin{appendix}

\section{Easy-to-use software for time delay estimation in presence of microlensing}
\label{sec:codes}

We consider a double quasar with images A and B. The A image magnitudes at observing times 
$t_i$ ($i = 1,\dots,N_{\rm A}$) are given by $A_i = m_{\rm A}(t_i)$, whereas $B_j = m_{\rm 
B}(t_j)$ at observing times $t_j$ ($j = 1,\dots,N_{\rm B}$) represent the light curve of B. 
The A image data are shifted by a time lag $\tau$ and a microlensing polynomial of degree 
$N_{\rm{ml}}$, and then compared to the B image data. The microlensing polynomial 
\begin{equation}
P(t_i) = a_0 + a_1 t_i + a_2 t_i^2 + ... + a_{N_{\rm{ml}}} t_i^{N_{\rm{ml}}}
\end{equation}
is characterised by coefficients $a_k$ ($k = 0,\dots,N_{\rm{ml}}$), and the central idea is 
to find the values of ($\tau$, $a_k$) that provide the best match between the shifted light 
curve of A and the original light curve of B.

To estimate best values of ($\tau$, $a_k$), we minimise two different merit functions: 
dispersion $D^2_4$ and reduced chi-square $\chi^2_{\rm r}$ (see main text and here below). 
Two main advantages of using a polynomial function to account for the microlensing 
variability are the possibility to describe a very general behaviour and the linear role of 
polynomial coefficients in merit functions. Thus, for each value of $\tau$, we can optimise 
the linear parameters $a_k$ analytically, so multivariable merit functions are reduced to 1D 
spectra depending on the non-linear parameter $\tau$. Minima of these 1D spectra yield 
estimates of the time delay between A and B. We also note that Python scripts to use both 
techniques (merit functions) are available at the GitHub repository URL 
https://github.com/glendama/q0951time\_delay, and draw attention to the fact that the 
so-called polynomial order in the scripts (Nord) equals $N_{\rm{ml}} + 1$.
  
\subsection{Minimum dispersion method}

The merit function is given by
\begin{equation}
D^2_4(\tau, a_k) = \frac{\sum_i \sum_j W_{ij} S_{ij}[A_i - B_j + P(t_i)]^2}{\sum_i \sum_j 
W_{ij} S_{ij}} ,
\end{equation}
where $W_{ij} = 1/(\sigma_{A_i}^2 + \sigma_{B_j}^2)$ are statistical weights computed from 
photometric errors and $S_{ij} = \exp[-(t_i - t_j + \tau)^2/\beta^2]$ are weights for time 
separations. Our time-separation weighting scheme relies on a Gaussian function and a  
"decorrelation length" $\beta$, and it is inspired by Eq. (13) of 
\citet{1996A&A...305...97P}. 

For each value of $\tau$, the minimisation of the dispersion with respect to the polynomial 
coefficients, i.e., $\partial D^2_4/\partial a_k = 0$, produces the system of linear 
equations 
\begin{equation}
\sum_i \sum_j W_{ij} S_{ij} \sum_l t_i^{l+k} a_l = -\sum_i \sum_j W_{ij} S_{ij} (A_i - B_j) 
t_i^k .
\end{equation}
This system of $N_{\rm{ml}} + 1$ equations can be solved by standard procedures of linear 
algebra, yielding a set of optimal microlensing coefficients $a_k$ for each time lag and the
corresponding dispersion spectrum $D^2_4(\tau)$. 

\subsection{Reduced chi-square minimisation}

The merit function has the form
\begin{equation} 
\chi^2_{\rm r}(\tau, a_k) = \frac{1}{N_{\rm{pair}} - (N_{\rm{ml}} + 2)} \sum_i W_i [A_i - 
B'_i + P(t_i)]^2 ,
\end{equation}
where $B'_i$ is the B image magnitude binned around $t_i + \tau$ to build the $(A_i, B'_i)$ 
pair, $N_{\rm{pair}}$ is the total number of $AB'$ pairs for the time lag $\tau$ ($i = 
1,\dots,N_{\rm{pair}}$), and $W_i = 1/(\sigma_{A_i}^2+\sigma_{B'_i}^2)$ is the inverse of 
the sum of squared uncertainties. To build the bin in B around $t_i + \tau$, we use a linear 
weighting function $1 - |t_i - t_j + \tau|/\alpha$ for dates $t_j$ verifying $|t_i - t_j + 
\tau| \leq \alpha$. 

The minimisation with respect to the microlensing coefficients $a_k$ yields the linear 
system 
\begin{equation}
\sum_i W_i \sum_l t_i^{l+k} a_l = -\sum_i W_i (A_i - B'_i) t_i^k ,
\end{equation}
and thus optimal values of $a_k$ for each time lag and the 1D spectrum $\chi^2_{\rm 
r}(\tau)$. 

\end{appendix}

\end{document}